# An Insight in Explaining the Stress Distribution in and around EGS

Mahmood Arshad, Masami Nakagawa, Kamran Jahanbakhsh, and Lucila Dunnington

Department of Mining Engineering, Colorado School of Mines, 1600 Illinois St Golden CO 80401

smahmoodarshad@yahoo.com

## ABSTRACT

Developing an Enhanced Geothermal System, otherwise known as EGS, is a complex process and is dependent on range of geological and operating variables. Stresses in and around EGS are believed to be either ‘sound and non-harming’ or ‘violent and catastrophic’ among different groups involved, directly or indirectly, in EGS. Pros and cons of EGS have been under discussion and research for a while now. This paper addresses issues associated with stress redistribution during and after working cycle of EGS and gives a new insight in understanding the behavior of stresses redistribution in and around basement rock. As the basement rock is thermo-elastically connected to the country rock, newly generated stresses interact with the existing in-situ stresses under prevailing conditions of geological, design and operating variables. Variables dictating the continuous, safe and efficient working of EGS are also outlined in detailed sections. Guidelines for future of the research related to stress redistribution in EGS are also part of this paper.

## 1. INTRODUCTION

Enhanced Geothermal Systems (EGS) is one of innovative energy winning technologies inspired by the concept of green and renewable energy, and has been under the radar of scientific and engineering researchers for almost 40 years. As EGS is still a developing technology and a complex phenomenon involving a range of geological and operating variables, there are many unanswered questions about how EGS works and what changes take place within and surrounding the EGS reservoir, during and after working cycle of EGS. This paper provides an explanation of heat flow and stress related behavior of EGS and variables dictating the continuous, safe and efficient working of EGS. Future research related to stress redistribution in EGS are also part of this paper.

## 2. EGS, HEAT FLOWS AND STAKES INVOLVED

EGS, also referred to as HWR (Hot Wet Rock), HDR (Hot Dry Rock) or HSR (Hot Sedimentary Rock) systems, is essentially a subsurface heat exchange system in which a network of fractured rock at a certain depth with suitable temperatures is stimulated and a fluid flow is established through injection and production borehole(s) to recover the earth’s heat to be used in further applications. Heat in EGS is sourced by a combination of natural radioactivity, earth’s heat of formation and combined flow of heat by conduction, advection and radiative transport (R. Jeanloz, 2013). Even for a small scale EGS system, radioactive heat production will require millions of years to produce enough temperatures to be useful for EGS, so earth’s radioactivity as a source of heat for EGS is not significant (Furlong, et al., 2013). Flow of heat for distances over tens of kilometers by conduction is also insignificant even for earth’s thermal diffusion constant of 1 $mm^2/s$. Radiative transport of heat is negligible for the earth’s physical state (Donald L. Turcotte, 1982). Most effective source for a hot crust is by advection involving earth’s heat of formation (Stolpher, et al., 1981). EGS is just one of the many ways of using Earth’s heat or geothermal energy for a range of applications. The current capacity of electrical generation using geothermal energy sources in US with a capacity factor of 70% is around 3.4 GW and is estimated to be 9 GW by USGS in 2008 with the development of known geothermal resources (R. Jeanloz, 2013).

DOE is planning to fully implement its Frontier Observatory for Research in Geothermal Energy, commonly known as FORGE, by 2020 (Lester, 2015) and is currently focusing on five different active EGS demonstration projects in four different states including Nevada, Idaho, Oregon and California (DOE, 2015). Although, water regulation authorities in different states of the United States have different water management regulations, water boards and Environmental Protection Agencies (EPA) have not shown any major reservations regarding this breakthrough technology (Cichon, 2013). Costs of power generation from EGS tend to be higher in early startup days but drops down rapidly as the system progresses in action, ranging from USD 2,500 to 5,000 per installed kW and USD 0.01 to 0.03 per KW as maintenance and operation costs (DOE, 2015). Power generation is not the only proposed use of energy recovered by Enhanced Geothermal Systems. Many other industrial (agriculture, aquaculture, greenhouse, drying, and process heat) and domestic uses (space heating, snow melting, and bathing and swimming parlors) have taken advantage of heat from earth. Cascading use of recovered heat is also one of the available options for fully utilizing what EGS has to offer (Lund, 2010).

## 3. ISSUES WITH EGS

There are some issues that need monitoring and remedial measures in order for EGS to function properly for the designed span of life. Flow through fractures and heat removal may cause dilation of fracture apertures and, as fluid flows through the path of least resistance, it can lead to fluid short circuiting (Ghassemi, et al., 2008) (Ghassemi, et al., 2007). In such a case, EGS may not be able to recover the estimated amount of heat.

Chemical or mechanical alteration of basement rock will occur as a result of chemically active or continuous fluid flow. Although water is supposedly only fluid medium of EGS flow network and fluid chemistry and flow rates may be designed in accordance with the properties of base rock (MIT, 2006).

EGS is considered responsible for induced seismicity during its development and production phase. Each fracture stimulating techniques used in EGS such as hydrofracturing, fluid injection and acidization has reportedly contributed to induced seismicity but predominantly only microearthquakes (Majer, et al., 2007) (MIT, 2006).

Heat recovery from earth's crust, being the primary objective of EGS, may be responsible for the changes in big picture of the project. All EGS factors, directly or indirectly related to EGS operation, ultimately play their role towards changes in stress in and around the basement rocks.

## 4. STRESSES CHANGES INVOLVED IN EGS

### 4.1 Basic Hypothesis

EGS reservoir, itself, is a subject that attracts most of scientific and research work but the bigger picture outside the EGS reservoir is the focus here. The question is, what happens outside the EGS reservoir when heat flows start to cause stress redistribution in the country rock during and after working life of EGS. The basic assumption, in answering this question, is that the country rock possess fair degree of homogeneity in general.

Once EGS is setup, that is, inlet and outlet boreholes drilled, fractures stimulated, flow through fractures established, and heat recovery process started, the basement rock is subjected to slow but continuous thermal and, in turn, structural changes. As the geology over different areas and depths is factually different and EGS is, ideally, supposed to be setup anywhere with supporting ground conditions, there isn't much to say with absolute certainty about what's going to happen regarding new stress state of the area. But knowing the basic behavior of the rock in reaction to thermal changes and flow through fractures, one can have an idea or predict of what's next in structural and stress transition.

Naturally occurring stresses in the earth simply relates to height of overburden and after a certain depth, vertical and lateral stresses are in unit ratio. Cooling down of the rocks causes them to shrink. The rocks at usual EGS depths (3-5 km or ~10,000 to 16,000 ft) are thought to exhibit thermoelastic behavior. This is why this shrinking as a result of cooling down generates tensile stresses in the surrounding rock. Newly generated tensile stresses interact with the existing state of stress of develop a whole new pattern of stresses in the EGS zone. It is the new stress pattern that decides next resulting events. Slight variation in the resulting stress state can define the mode, time and occurrence of next impactful event and one event can lead to another. This whole chain or cycle of EGS process during its full life, is illustrated in figure 1 and depends on different sets of variables involved in EGS.

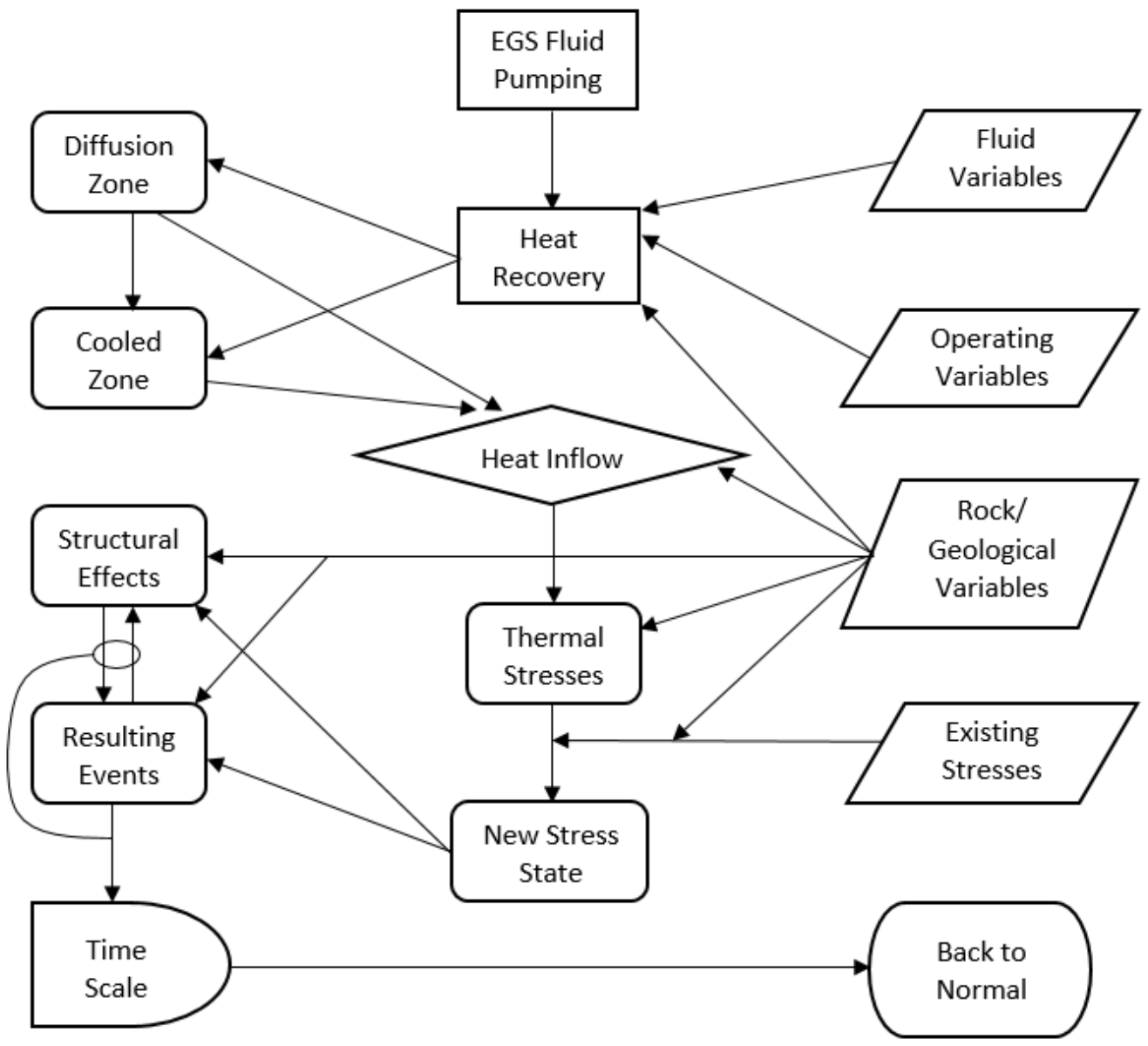

**Figure 1: EGS flow chart.**

Bottom line of the scenario is that the in-situ stress state is going to go under changes. Estimating the span of the changes, timeline of occurrence of eventual events, magnitude of such changes and extent to which it can affect the surrounding settings is not a simple deal (Khademian, et al., 2012). For an efficient EGS operation, there must be a balance between heat outflow (heat recovery) and heat inflow towards the basement rock. This means a continuous heat flow must exist from the country rock towards basement rock. Heat flows in the EGS lead to occurrence of different thermal zones in the area. Thermal zones and cooling fronts do not just exist along a straight line between inlets and outlets but they exist and expand in the vicinity, in more like a distorted spherical shape. Figure 2 illustrates different thermal zones during the EGS cycle. These thermal zones have been explained in DOE report on EGS (R. Jeanloz, 2013). Diffusion front separates the rock still at ambient temperature from diffusion zone, which is where rocks start to cool down and the temperature is still between ambient temperature and injection water temperature. Cooling front separates the diffusion zone and cooled zone where the rock temperature has dropped down to injection fluid temperature and doesn't add much heat to water.

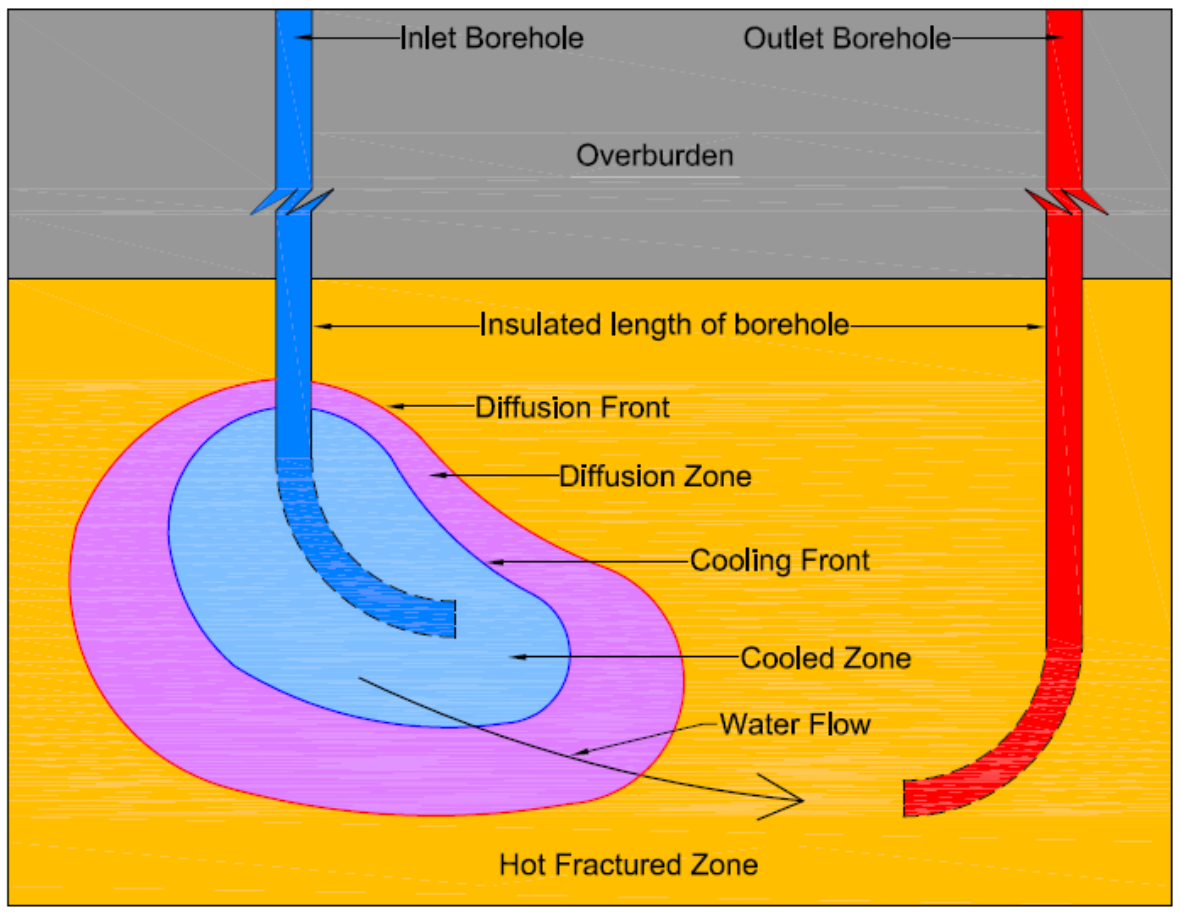

**Figure 2: Heat zones in EGS basement rock.**

There are two basic facts associated with heat flows in EGS. First, a continuous heat inflow exists from the surrounding to EGS basement rock. Second, there exists earth's natural thermal gradient, implying rock temperatures above the basement is already less than the temperatures being dealt with. These facts suggest that most of the heat inflows are from the rocks in all directions but from the rock above the EGS basement rock. This doesn't mean that there will be absolutely no inflow from upper level; it will be just low inflow compared to other directions or negligible inflow. To elaborate more, imagine the basement rock is a perfect cube with six sides; upper side is negligible heat inflow. Four sides, perpendicular to horizontal plane, have ideally same temperature and heat flow conditions in the surrounding. Bottom side of the cube has highest temperature difference for heat flow but rest of the heat flow conditions are still same. There are two sides for vertical heat inflow; the fact that upper side provides negligible heat inflow, leaves only one side for vertical heat inflow. On the other hand, there are four sides available for horizontal heat inflow ideally providing similar heat inflows. This enlightens a fact that most of the heat inflows, thermal changes and its stress and structural effects have to occur in lateral direction and in the rock under EGS basement; more in the surrounding of inlet borehole compared to the outlet. Realizing the fact that there isn't any perfect cube and distorted spheres and curves are actually being dealt, gives more of a fat penny shaped volume for heat inflows, similar to penny shape concept used in petroleum industry. Figure 3(a) is the illustrative explanation of the fact and figure 3(b) is the illustration of what ultimate diffusion or ultimately, impact zone may look like.

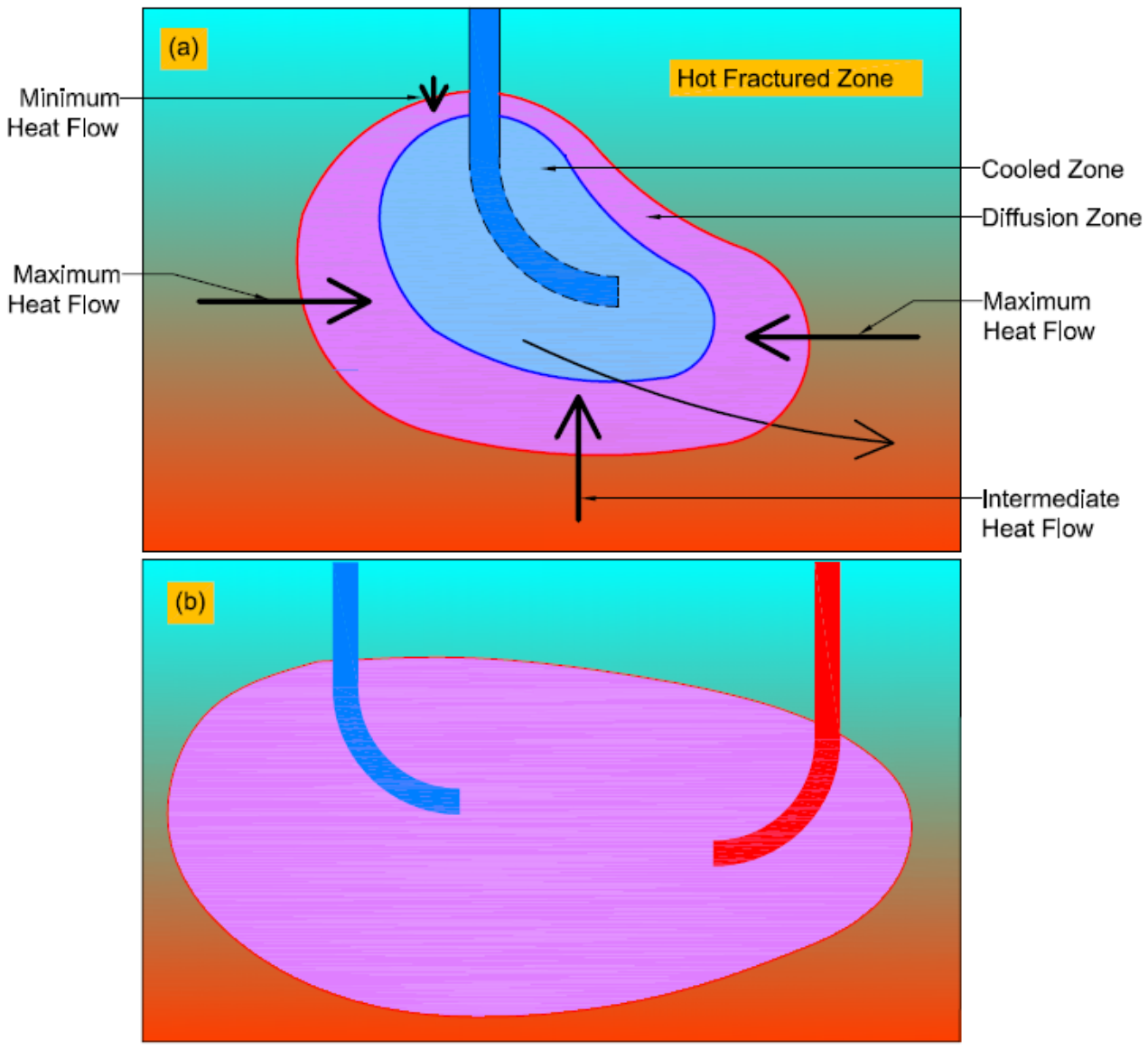

**Figure 3: (a) Heat flows in EGS basement rock (b) Ultimate diffusion/impact zone.**

Occurrence of diffusion or ultimately, impact zone more in lateral plane suggests the impacts of the EGS to be majorly in lateral dimension. Another factor to be considered here is the timeline of the whole process, considering the life of EGS to be 25-50 years, it is going to take even longer, with earth's thermal diffusion constant, for the impact zone to nullify the temperature changes and regain the ambient temperature. Such a slow and time taking process that supposedly has major impacts laterally, may be suggested to accommodate our energy needs in future.

### 4.2 Variables involved in EGS:

There are many variables involved in defining the final stress state around EGS, and any combination of these variables may play role in a given EGS scenario. These variables may include subsets of geology of the area, rock physical and chemical properties, fluid and flow attributes, and design parameters. These variables are outlined below.

| **Fluid Related** | **Operating Variables** | **Rock/ Geological Variables** |
|---|---|---|
| Fluid viscosity<br>Specific Heat<br>Fluid Compressibility<br>Temperature Points | Fluid flow rate<br>Rate of heat removal<br>Pattern and number of injection and production boreholes<br>Chemical alteration<br>Fatigue/ mechanical breakage | Porosity and Permeability<br>Thermal conductivity<br>Heat flux<br>Stratigraphy<br>Constituent Composition<br>Heterogeneity<br>Porosity and permeability of rock<br>Fracture pattern: size, span, width, distribution, active or inactive<br>Pore pressure<br>Pore volume |

## 5. CONCLUSION/ REMARKS

To summarize, the basic points of hypothesis are as follows:

- When an EGS system is put in action, it recovers heat from the basement rock.
- Heat removal from basement rocks initiates a net heat inflow from the surrounding rock to the basement rock.
- Continued heat flows create different thermal zones in the area.
- Net cooling of the rock causes rock to shrink.
- Shrinking rocks generates tensile stresses in the surrounding of the basement rock.
- Generated tensile stresses interact with existing stresses to create a whole new stress pattern.
- New stress pattern impacts and affects the surrounding depending on operating, flow, and geological variables and timeline of the system.
- The impacts and effects may occur in a chain or cyclic manner.
- Ultimate impact zone is suggested to exist in a fat penny shape.
- Major changes are believed to occur in lateral direction.
- This slow and time taking process that supposedly has major impacts laterally, with the assumption of general homogeneity in the surrounding, may be suggested to fulfil our future energy needs.

## 6. FUTURE RESEARCH

Validating this theory and quantifying involved stresses and their impacts, through physical modeling and, if possible, field results of different ongoing projects is going to be the starting point. Studying the behavior and effects of stress redistribution due to thermal changes in the surrounding of EGS reservoir is mainly the aim of this research project. What happens outside the reservoir and time, extent and cyclic behavior of the impacts, depending on the EGS variables, is the focus of my research. Another interesting question related to this research is looking into the possibilities without the basic assumption of this proposed theory. That is, what would be the scenario if the surrounding of EGS reservoir is non non-homogenous in general. This theory may also be improved using experts' opinions and comments and an improved version may be presented on a later date.